\documentclass[sigconf]{acmart}
\usepackage{newfloat}
\usepackage{listings}
\usepackage{booktabs}
\usepackage{mathtools}
\usepackage{multirow}
\usepackage{bbm}
\usepackage{algorithm}
\usepackage{arydshln}
\usepackage{algpseudocode}
\usepackage{xcolor}
\usepackage{tikz}

\AtBeginDocument{%
  \providecommand\BibTeX{{%
    \normalfont B\kern-0.5em{\scshape i\kern-0.25em b}\kern-0.8em\TeX}}}

\copyrightyear{2024}
\acmYear{2024}
\setcopyright{acmlicensed}
\acmConference[SIGIR '24]{Proceedings of the 47th International ACM SIGIR Conference on Research and Development in Information Retrieval}{July 14--18, 2024}{Washington, DC, USA}
\acmBooktitle{Proceedings of the 47th International ACM SIGIR Conference on Research and Development in Information Retrieval (SIGIR '24), July 14--18, 2024, Washington, DC, USA}
\acmDOI{10.1145/3626772.3657797}
\acmISBN{979-8-4007-0431-4/24/07}

\begin{document}

\title{Generative Retrieval via Term Set Generation}

\author{Peitian Zhang}
\authornote{Co-first author.}
\orcid{0009-0007-1926-7433}
\email{namespace.pt@gmail.com}
\affiliation{%
  \department{Gaoling School of Artificial Intelligence}
  \institution{Renmin University of China}
  \city{Beijing}
  \country{China}
}

\author{Zheng Liu}
\authornotemark[1]
\authornote{\vspace{-4.8pt}Corresponding author.}
\email{zhengliu1026@gmail.com}
\affiliation{%
  \institution{Beijing Academy of Artificial Intelligence}
  \city{Beijing}
  \country{China}
}

\author{Yujia Zhou}
\affiliation{%
  \department{Gaoling School of Artificial Intelligence}
  \institution{Renmin University of China}
  \city{Beijing}
  \country{China}
}

\author{Zhicheng Dou}
\authornotemark[2]
% \authornote{Corresponding author.}
\email{dou@ruc.edu.cn}
\affiliation{%
  \department{Gaoling School of Artificial Intelligence}
  \institution{Renmin University of China}
  \city{Beijing}
  \country{China}
}

\author{Fangchao Liu}
\author{Zhao Cao}
\affiliation{%
  \institution{Huawei Poisson Lab}
  \city{Beijing}
  \country{China}
}


\renewcommand{\shortauthors}{Zhang, et al.}
\newcommand{\red}[1]{\textcolor{red}{#1}}
\newcommand{\blue}[1]{\textcolor{blue}{#1}}
\newcommand{\cyan}[1]{\textcolor{cyan}{#1}}
\newcommand{\green}[1]{\textcolor{green}{#1}}
\algnewcommand{\LineComment}[1]{\State\textcolor{blue}{\(\triangleright\)#1}}
\newcommand{\highlight}[2][yellow!50]{%
  \tikz[baseline=(X.base)]{
    \node[fill=#1, inner sep=1pt] (X) {#2};
  }%
}


%\begin{CCSXML}
%<ccs2012>
%   <concept>
%       <concept_id>10002951.10003317.10003338.10003341</concept_id>
%       <concept_desc>Information systems~Language models</concept_desc>
%       <concept_significance>500</concept_significance>
%       </concept>
%   <concept>
%       <concept_id>10002951.10003317.10003338.10010403</concept_id>
%       <concept_desc>Information systems~Novelty in information %retrieval</concept_desc>
%       <concept_significance>500</concept_significance>
%       </concept>
%\end{CCSXML}
%
%\ccsdesc[500]{Information systems~Language models}
%\ccsdesc[500]{Information systems~Novelty in information retrieval}



\begin{abstract}
Recently, generative retrieval has emerged as a promising alternative to the traditional retrieval paradigms. It assigns each document a unique identifier, known as the DocID, and employs a generative model to directly generate the relevant DocID for the input query. A common choice for the DocID is one or several natural language sequences, e.g. the title, synthetic queries, or n-grams, so that the pre-trained knowledge of the generative model can be effectively utilized. However, a sequence is generated token by token, where only the most likely candidates are kept and the rest are pruned at each decoding step, thus, retrieval fails if any token within the relevant DocID is falsely pruned. What's worse, during decoding, the model can only perceive preceding tokens in the DocID while being blind to subsequent ones, hence is prone to make such errors. To address this problem, we present a novel framework for generative retrieval, dubbed \textbf{T}erm-\textbf{S}et \textbf{Gen}eration (TSGen). Instead of sequences, we use a set of terms as the DocID. The terms are selected based on learned weights from relevance signals, so that they concisely summarize the document's semantics and distinguish it from others. On top of the term-set DocID, we propose a permutation-invariant decoding algorithm, with which the term set can be generated in any permutation yet will always lead to the corresponding document. Remarkably, TSGen perceives all valid terms rather than only the preceding ones at each decoding step. Given the constant decoding space, it can make more reliable decisions due to the broader perspective. TSGen is also resilient to errors: the relevant DocID will not be falsely pruned as long as the decoded term belongs to it. Moreover, TSGen can explore the optimal decoding permutation of the term set on its own, which further improves the likelihood of generating the relevant DocID. Lastly, we design an iterative optimization procedure to incentivize the model to generate the relevant term set in its favorable permutation. We conduct extensive experiments on popular benchmarks of generative retrieval, which validate the effectiveness, the generalizability, the scalability, and the efficiency of TSGen.
\end{abstract}



\maketitle

\vspace{-5pt}
\section{Introduction}\label{sec:intro}

Document retrieval, standing as the most representative form of information retrieval, is fundamentally important to real-world applications like web search, question answering, advertising, and recommendation~\cite{DPR-Karpukhin,IR_Survey-Zhao}. 
Nowadays, they are also regarded as a critical tool for the augmentation of large language models (LLMs), where external information can be introduced to enhance their knowledge and capability~\cite{WebGPT-Nakano,Shall_We_Pretrain_Autoregressive_Language_Models_with_Retrieval-Wang,zhang2023llmembedder}. 
Typical document retrieval methodologies call for two basic modules: representation and indexing. 
For example, a sparse retrieval system uses lexical representations and an inverted index~\cite{UniCOIL-Lin,BM25-Robertson}, while a dense retrieval system is based on dense embeddings and an ANN index~\cite{HNSW-Malkov,zhang2023hybrid}. 

Recently, generative retrieval~\cite{DSI-Tay,GENRE-Cao,NCI-Wang,Ultron-Zhou} emerges as a promising alternative to the traditional sparse or dense retrieval~\cite{Rethinking_Search-Metzler}. 
Specifically, it assigns each document a unique identifier, known as the DocID, and employs a generative model to directly generate the DocID of the relevant document for the input query.
Compared with traditional retrieval methods, generative retrieval is end-to-end differentiable: the entire retrieval pipeline including representation and indexing is encapsulated into a generative model and hence can be optimized by the Seq2Seq learning~\cite{Rethinking_Search-Metzler,DSI-Tay}.
% 一大部分工作用natural language sequence
Defining appropriate DocID is fundamental to effective generative retrieval.
A large body of research utilizes one or several natural language sequences as the DocID, such as the title~\cite{GENRE-Cao}, n-grams~\cite{SEAL-Bevilacqua,li2023minder,li2023learning_to_rank_in_generative_retrieval}, or synthetic queries~\cite{DSI_QG-Zhuang,tang2023inspired_by_learning_strategy}. 
This is because modern generative language models (e.g., T5~\cite{T5-Raffel}, Llama~\cite{touvron2023llama-a}) are pre-trained on natural language, thus, employing a natural language based DocID may seamlessly inherit the pre-trained knowledge of these models~\cite{SEAL-Bevilacqua}. Besides, they are also more interpretable and generalizable than other alternatives like naive ID~\cite{How_Does_Generative_Retrieval_Scale_To_Millions_Of_Passages-Pradeep} and clustering based ID~\cite{NCI-Wang,TIGER-Rajput}. 

% 生成式模型token-by-token地生成，在每一步生成时，都会根据概率选取top-k的hypothesis，这就意味着有很多hypothesis将会被丢弃，且再也无法回溯，即相关文档无法被检索到一旦解码过程中任何一步将其对应的DocID hypothesis丢弃了。
% 这个现象在生成式检索中很常见
% 这个现象使得sequence DocID的顺序变得tricky，因为对于不同的查询，有些prefix的生成概率高，因此不会被prune；另一些prefix的生成概率不够高，因此会被prune，从而影响检索效果。
% 直觉上，false prune问题很常见，因为预测下一位token时模型只能基于其前序token进行判断，完全忽视后序token的信息，从而更容易犯错。
Unlike natural language generation, which allows flexible paraphrasing for the same meaning, generative retrieval poses a unique challenge: it must exactly generate the relevant DocID sequence.
The sequence is generated token by token. At each decoding step, the top-$B$ ($B$: the beam size) likely candidates are kept and the rest are pruned; consequently, the relevant DocID cannot be generated if any token within it is falsely pruned.
Intuitively, this \textbf{false pruning problem} is common and challenging to mitigate: the generative model can only perceive preceding tokens in the DocID (tokens next to the current decoding step), without access to the information in subsequent ones, hence is prone to make mistakes.
This limits the overall retrieval quality.

\begin{figure*}
    \centering
    \includegraphics[width=0.9\textwidth]{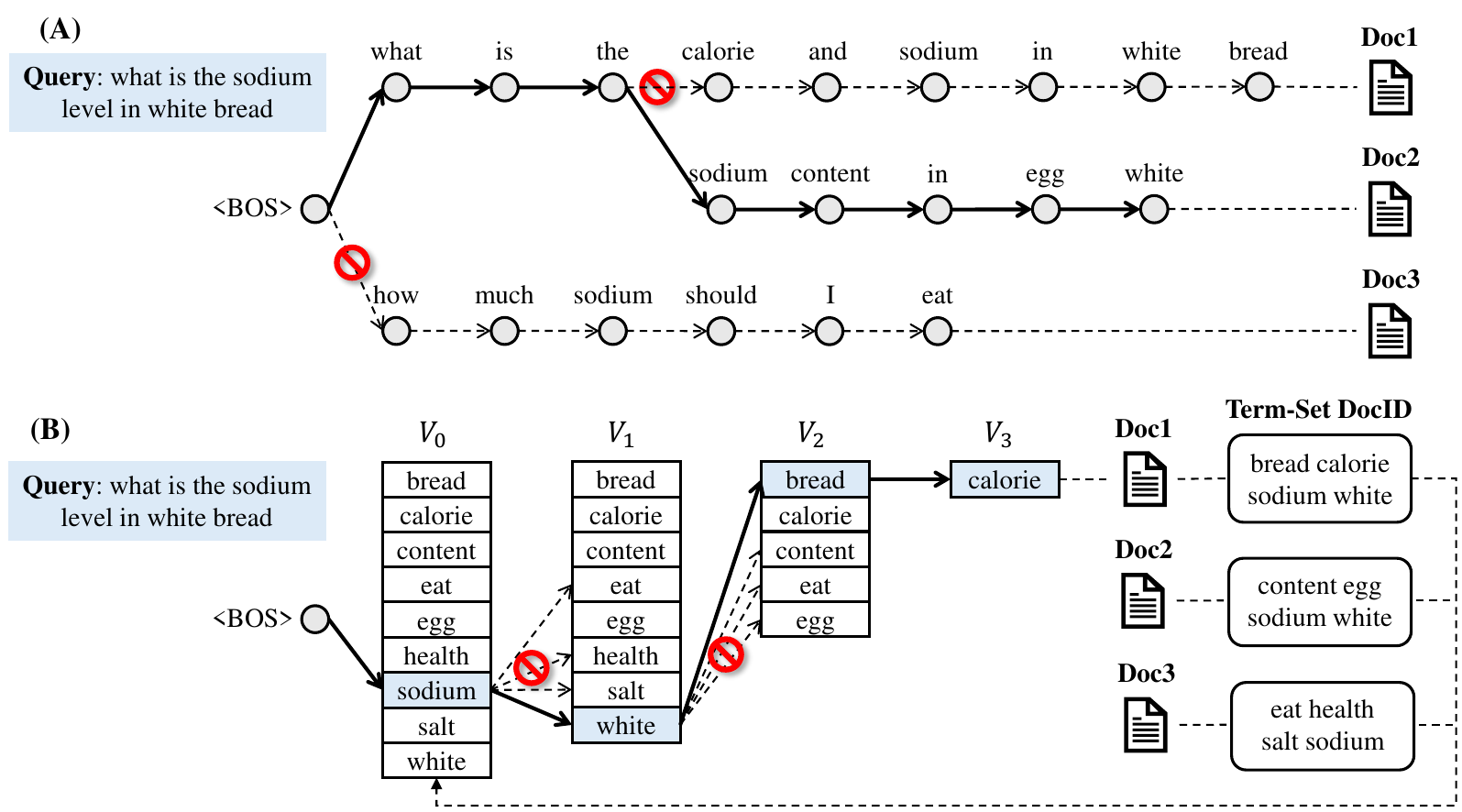}
    \vspace{-10pt}
    \caption{(1) An example of the false pruning problem stemming from the requirement of exactly generating the sequence DocID. The relevant document (Doc1) is falsely pruned during decoding due to the limited perspective of the model and the low resilience to errors.
    (2) TSGen successfully retrieves the relevant document in the same context. 
    Based on the term-set DocID and the permutation-invariant decoding, TSGen enjoys a broader perspective at each decoding step to make more reliable decisions and is resilient to errors.}
    \label{fig:TSGen}
    \vspace{-5pt}
\end{figure*}

% 用一个例子说明false pruning问题
% -> synthetic query DocID为例，如何泛化到别的natural language sequence DocID
% -> 更多sequence能否解决问题？
In Figure~\ref{fig:TSGen}(A), we introduce an example to better illustrate the false pruning problem with sequence DocID.
Without loss of generality, we use one query (\textit{``what is the sodium level in white bread''}) from the dev set of MSMARCO Dataset~\cite{MSMARCO-Nguyen} and consider three candidate documents where Doc1 is the relevant one. 
Following previous studies~\cite{tang2023inspired_by_learning_strategy}, we use one synthetic query as the DocID, and SE-DSI as the generative model. 
We can observe that the relevant DocID is falsely pruned in the 4-th decoding step. Specifically, the likelihood of its prefix is lower than that of the second DocID, i.e. $p(\text{what,is,the,calorie}\mid Q) < p(\text{what,is,the,sodium}\mid Q)$, even though the suffix of the second DocID (\textit{``egg white''}) is not relevant to the query at all.
As stated, this problem stems from the requirement for the exact generation of the relevant DocID sequence.
While existing approaches, such as increasing the beam size or maintaining multiple DocID sequences per document~\cite{SEAL-Bevilacqua,li2023minder}, provide some relief, they do not systematically solve the issue.

% 词集合作为docid
% 所有词加在一起全面且简练的总结了文档信息
% 每个词都很短，通常仅包含一个token，因此模型生成该词时能看到其全部信息，即不会因为忽视suffix而犯错
In this work, we present a novel framework, dubbed \textbf{T}erm \textbf{S}et \textbf{Gen}eration (TSGen), to address the above problem.
Instead of one or several sequences, TSGen uses a set of terms as the DocID.
These terms are selected based on the learned weights from relevance signals. As a result, they not only concisely summarize the document's semantics, but distinguish the document from others.

% decoding algorithm：词可以用任何顺序生成，但都指向到对应的文档
% 和constrained beam search一样都保证生成的docid一定是有效对应于一篇文档的
% 和term-set docid配合，带了了额外的优势
% 1. 因为term可以以任何顺序生成，在解码的每一步就多了N个容错，而解码总空间不变，从而进一步减轻了false pruning问题
% 2. 模型会根据查询自行选择最大化likelihood的permutation来生成，让相关文档更容易被找到
% 3. 对于没有见过的文档，由于其term有重合，模型能够将学习得到的相关关系泛化
On top of the term-set DocID, we propose a permutation-invariant decoding algorithm, with which the terms can be generated in any permutation yet will always lead to the corresponding document.
Akin to the widely used constrained beam search~\cite{GENRE-Cao,Ultron-Zhou}, the proposed algorithm guarantees the validity of the generated DocID.
In other words, the generated term set always corresponds to a valid document in the corpus.
In addition, it introduces three key advantages. 
(1) At each decoding step, TSGen perceives all valid terms rather than only the preceding ones, thereby acquiring full information about the DocID. 
Each term itself usually comprises only one token, thus the decoding space remains unchanged, and TSGen can make more reliable decisions given its broader perspective.
(2) TSGen is resilient to decoding errors. In contrast to sequence-based generation, the relevant DocID will not be falsely pruned as long as the decoded terms belong to it.
(3) TSGen can explore the optimal permutation on its own to generate the term set. 
It may permute the terms in one way for certain queries while in another way for others to maximize the likelihood.
This flexibility makes it easier to generate the relevant DocID and improves the retrieval accuracy.
The algorithm is implemented with an inverted index and is competitive in efficiency as elaborated in~Section~\ref{subsec:decoding}.

% 回到例子：每个文档使用term-set docid
% 生成时，模型每一步都在valid term space中选择概率最大的进行解码，其最初是所有term的并集，并随着解码过程一步步缩减从而保证validity
% 模型依次从V_i中解码出了sodium，white，bread，calorie，从而成功找到相关文档。Doc2和Doc3分别在第二和第三步被prune，然而这个prune依靠了其docid的全部信息（all terms），因此更加可靠。
% Notably，即使模型在第二位选错，比如解码到bread，Doc1仍然不会被prune，因为bread也包含在term set之中，这意味着模型false prune的chance变低了。
% 同时，对于不同的查询，模型可以自行探索最优permutation，比如"what is the calorie in white bread"，则模型倾向于生成“calorie，white，bread，sodium”
% 最后，对于新的关于煎鸡蛋放多少盐的文档，模型也能够将现有的knowledge应用到其上
Back to our example in Figure~\ref{fig:TSGen}(B), the three documents are paired with their term-set DocID. 
At the $i$-th decoding step, the model decodes the next term from the valid decoding space ($V_i$). 
Initially, this space encompasses the union of all terms, and it gradually narrows down during the decoding process to ensure validity.
We can observe that the model sequentially decodes \textit{``sodium, white, bread, calorie''} from $V_i$, successfully generating the relevant DocID.
% At each decoding step, the model is exposed to all valid terms, which removes the barrier that only the prefix can be seen.
At each decoding step, the model is exposed to all valid terms in all documents, so it can thoroughly examine all information in the DocID before making a decision. 
Besides, the model is allowed to use a different permutation than the highlighted one, for example, it may decode \textit{``bread''} or \textit{``calorie''} other than \textit{``white''} in the second generation step, which is just a different permutation of the term set so the relevant DocID can still be generated.
This error resilience applies to the pruning of negative documents as well.
In the second and third steps, Doc3 and Doc2 are pruned when all the terms of their DocID are considered irrelevant.
Moreover, the model can adjust the permutation of the term set for different queries. 
For instance, it may generate \textit{``calorie white bread sodium''} in response to the query \textit{``how much calories is in the white bread''}.

Finally, we devise an iterative optimization procedure to guide TSGen to generate the relevant term set in its favorable permutation. 
We employ the standard Seq2Seq loss while periodically update the permutation of the term set based on the latest model snapshot.
The model converges in a state where it can successfully generate the term set in the permutation that it reckons the most probable.

The contributions of our work are summarized as follows:

\noindent(1) We propose TSGen, a novel generative retrieval framework where a set of terms is used as the DocID, and any permutations of these terms lead to the corresponding document. 
It systematically addresses the false pruning problem of the widely used sequence-based DocID designs in existing generative retrieval methods.
    
\noindent(2) We devise tailored techniques for TSGen, including the learned term selection, the permutation-invariant decoding algorithm, and the iterative optimization procedure.

\noindent(3) We conduct extensive experiments on popular generative retrieval benchmarks. The results validate the effectiveness, generalizability, scalability, and efficiency of TSGen.

\vspace{-5pt}
\section{Related Work}
\subsection{Traditional Document Retrieval}
Document retrieval has been extensively studied for a long time. There are mainly two threads of research towards traditional document retrieval. 
One of them is \textit{sparse retrieval}, which stores the bag-of-words (BOW) representation of a document and estimates relevance based on the weights of overlapping terms, exemplified by BM25 algorithm~\cite{BM25-Robertson}. 
Another one is \textit{dense retrieval}, which fine-tunes a pretrained language model to encode the document and the query in a low-dimensional semantic space, and model the retrieval problem as the maximum inner product search (MIPS)~\cite{DPR-Karpukhin,bge_embedding,bge_m3,llm_embedder}. 
When applied with tailored training techniques, including mining hard negatives~\cite{ANCE-Xiong,Star-Zhan,RocketQA-Qu}, distillation~\cite{AR2-Zhang,ERNIE_Search-Lu}, and retrieval-oriented pretraining~\cite{RetroMAE-Xiao,SimLM-Wang}, dense retrieval has achieved state-of-the-art effectiveness.
In the meantime, continuous efforts have been made to enhance the sparse retrieval with PLMs to learn contextualized term weights~\cite{DeepImpact-Mallia,DeepCT-Dai,UniCOIL-Lin,SPLADE-Formal,SPLADEv2-Formal}. 
All the aforementioned retrieval methods rely on standalone indexes to work, such as the inverted index~\cite{Inverted_Index-Zobel} and the ANN index~\cite{Faiss-Johnson}. These separate indexes cannot be optimized together with the encoding model. 

\vspace{-5pt}
\subsection{Generative Retrieval}
Recently, generative retrieval is proposed as an alternative to the traditional index-then-retrieve pipeline~\cite{DSI-Tay}, where a generative model (usually a Seq2Seq language model like T5~\cite{T5-Raffel}) is fine-tuned to directly map the input query to its relevant document. 
In this formulation, the document is represented by its identifier, namely the DocID, a much shorter signature than the document itself and hence easier to generate. 
Meanwhile, the index is replaced with the parametric memory of the language model and hence can be optimized end-to-end. 
As a result, the DocID becomes one of the most decisive factors for generative retrieval: the model must generate the exact same DocID for the targeted document, and the ranking of the document is determined by the generation likelihood of its DocID. 
Based on the choices of DocIDs, current works can be roughly partitioned into four groups: 

\noindent(1) The atomic ID based methods~\cite{DSI-Tay,Ultron-Zhou,DSI_QG-Zhuang,dynamic_retriever} use an integer or a tokenized integer. They lack semantics since there's no relation between the document's content and its DocID. Consequently, it requires the model to remember all DocIDs during training.

\noindent(2) The clustering based methods~\cite{DSI-Tay,DSI++-Mehta,NCI-Wang,CLEVER-Chen} use a series of cluster center indices. 
Prevalent approaches include hierarchical KMeans~\cite{DSI-Tay,NCI-Wang}, product quantization~\cite{Ultron-Zhou,CLEVER-Chen}, and residual quantization~\cite{TIGER-Rajput,IRGen-Zhang,zeng2023scalable_and_effective}. 
Despite their improvements, these DocIDs are not directly compatible with pre-trained generative model because their embeddings in the decoding table are newly introduced.

\noindent(3) The learned DocID based methods~\cite{yang2023auto_search_indexer,sun2023learning_to_tokenize} progressively learn the DocID based on document reconstruction or relevance signals.
Similar to the clustering based ones, these DocIDs must be optimized apart from the pre-trained weight of the generative model.

\noindent(4) The natural language based methods~\cite{GENRE-Cao,de-cao-etal-2022-multilingual,SEAL-Bevilacqua,Ultron-Zhou,CorpusBrain-Chen,UGR-Chen,tang2023inspired_by_learning_strategy,uni_gen} use one or several natural language sequences, e.g., the URL, the title, n-grams, or synthetic queries. 
These methods may directly inherit the enhanced capability from the pre-trained generative model. Besides, they are more interpretable than other alternatives.
However, we argue that the natural language sequences naturally suffer from the false pruning problem. 
As a remedy, our work adopts a set of terms as DocID, systematically mitigating the false pruning problem. 

\vspace{-5pt}
\subsection{Set Generation}
The concept of set generation originates from the Set Prediction Networks~\cite{Deep_Set_Prediction_Networks-Zhang}, which optimizes permutation-invariant losses to learn to decode a set of vectors based on the input vector. There have also been successful implementations to generate a set of texts using transformers~\cite{Stochastic_Optimization_of_Text_Set_Generation-Hashemi}. However, instead of generating independent elements one by one, we seek to generate a term set in its optimal permutation. 
This is somehow relevant to the autoregressive graph generation, which generates a graph by sequentially adding nodes and edges~\cite{GraphRNN-Jin,GRAN-Liao,GraphGEN-Goyal,Fitting_Autoregressive_Graph_Generative_Models-Han}. 
Multiple autoregressive node-edge sequences (intractable) may lead to the same graph, yet only a few of them are believed to be easier for model learning since they may follow some straightforward patterns. 
% Different heuristics and approximations are adopted to explore the optimal generation order for the learning of better models, e.g. using a subset of all possible orders or leveraging variational inference to marginalize all orders. 
By comparison, our task is even more challenging as there are no given nodes and edges. In other words, we have to construct the term set from scratch and then determine its generation order. In this paper, we propose the learned term selection to extract discriminative terms, and the iterative optimization to guide the model to generate the relevant term set in its favorable permutation.

\section{Methodology}
% 定义生成式检索问题：docid、解码算法、优化策略
In generative retrieval, each document is associated with its identifier, namely the DocID.
A generative model is employed to generate the DocID of the relevant document for the query. 
The relevance between the query $Q$ and the document $D$ can be expressed as:
\begin{equation}
\mathrm{Rel}(Q, D) = \Psi\left(\left\{\prod\nolimits_{i=1}^{|I|} p(I_i\mid I_{<i},Q;\Theta):~ I\in\mathcal{I}(D)\right\}\right), 
\end{equation}
where $\mathcal{I}(D)$ denotes all the DocIDs assigned to $D$ and $\Psi$ is a family of functions to aggregate multiple scalars into one.
Usually, there is only one DocID for each document, so $\Psi(\cdot)$ is the identity function $\mathbbm{1}(\cdot)$.
SEAL~\cite{SEAL-Bevilacqua} and its followers~\cite{UGR-Chen,li2023minder} assign multiple sequences to each document as the DocID. In that case, $\Psi(\cdot)$ is implemented with an intersective scoring function based on heuristics.

Computing $\mathrm{Rel}(Q,D)$ with all documents in the corpus is computationally prohibitive, thus, existing methods resort to constrained beam search for efficient generation. 
In brief, at each decoding step, only $B$ candidates are selected for a further generation while others are all pruned out based on the likelihood of their prefixes (i.e. $\sum_{i=1}^{*}\log p(I_i\mid I_{<i},Q;\Theta)$).
However, the generative model is likely to falsely prune the relevant docid by mistake since it can only perceive preceding tokens without access to subsequent ones.

In this work, we propose the Term Set Generation (TSGen) framework to address this problem. 
Specifically, instead of one or several sequences, TSGen uses a set of terms as the DocID, denoted as $\mathcal{T}(D)=\{\tau_1,\dots,\tau_N\}$.
These terms are automatically selected to concisely summarize the information in the document and distinguish it from others.
On top of the term-set DocID, we revolutionize the constrained beam search with a permutation-invariant decoding algorithm, with which any permutation of the term set always leads to the corresponding document\footnote{We explicitly guarantees the uniqueness of the term sets as described in $\S$~\ref{subsec:term_selection}, $\S$\ref{subsec:implementation}.}.
Meanwhile, the model can explore the optimal permutation of the term set on its own. This translates to an aggregation function of $\Psi(\cdot)\gets\max(\cdot)$.
Finally, the model is incentivized to generate the term set in its favorable permutation through an iterative optimization procedure.

In the remaining part of this section, we will elaborate on the learned term selection (Section~\ref{subsec:term_selection}), the permutation equivariant decoding (Section~\ref{subsec:decoding}), and the iterative optimization (Section~\ref{subsec:optimization}).

\subsection{Learned Term Selection}\label{subsec:term_selection}
% 词
The quality of the terms is critical to the retrieval quality of TSGen.
They should be informative because otherwise they do not bring about information and may interfere with the decoding process.
They should also be discriminative so that the model can better learn the relation between the query and the term sets.
In this place, inspired by existing sparse retrieval methods~\cite{UniCOIL-Lin,SPLADEv2-Formal}, we employ a term selection module to learn from the query-document relevance and perform accurate term selection.

Specifically, each document $D$ is first partitioned into a list of terms: $[t^D_1, \dots, t^D_{|D|}]$. 
Then, the terms are encoded into hidden states by a BERT. 
Finally, the weight of each term is obtained by pooling its hidden state into a scalar through an MLP layer (two linear transformations plus one dropout).
Formally,
\begin{align}
\label{eq:term_selection}
\boldsymbol{e}^D_{i} &= \mathrm{BERT}([t^D_{1}, \dots, t^D_{|D|}])[i], \quad 
w^D_{i} = \mathrm{MLP}(\boldsymbol{e}^D_{i}).
\end{align}

The same operation is performed on the query $Q$ as well, resulting in the term weights $\{w_j^Q\}_{j=1}^{|Q|}$.
Both $\mathrm{BERT}$ and $\mathrm{MLP}$ are learned with contrastive learning based on the relevance signals $\mathcal{A}=\{\langle Q,D^+, \{D_i^-\}_{m=1}^M \rangle \}$ where $D^+$ is the relevant document to $Q$, and $\{D_i^-\}_{m=1}^M$ are $M$ hard negatives mined from BM25:
\begin{gather}
    \mathcal{L}=\min\left(-\log \frac{\exp(s(Q,D^+))}{\exp(s(Q,D^+)) + \sum_{m=1}^M\exp(s(Q,D^-_m))}\right),\\
    s(Q,D) = \sum\nolimits_{t^Q_j = t^D_i} w^Q_j w^{D}_i
\end{gather}
where $t_j^Q=t_i^D$ indicates $t_j^Q$ and $t_i^D$ are the same term.

Therefore, the module learns to assign high weights to those terms that not only represent the whole documents, but also distinguish the relevant document from the irrelevant ones.
Based on these well-learned weights, we select the top-$N$ terms for each document to formulate the term-set DocID:
\begin{equation}
\mathcal{T}(D) = \left\{t_i^D: w^D_i \in\text{top-}N\left(\{w^D_i\}_{i=1}^{|D|}\right)\right\}.
\end{equation}
In practice, we choose the smallest value of $N$ while ensuring the uniqueness of the term set. For example, $N=12$ is already enough for a small-scale corpus like NQ100K.

% To keep the select term set informative, we first stem all terms by porter stemmer~\cite{Pyserini-Lin}, then deduplicate them, where the largest weight is kept for each unique term.
% Next, we select the top-$N$ terms by their weights.
% In our implementation, we find there are very few identifier collisions, i.e. the same set of keywords are selected for two different documents, under a moderate setting of $N$ (e.g. 12). We examine the collision cases and find the documents are actually two different editions of the same article, thus we keep them as is. It is worth noting that our method can be generalized to select phrases or any n-grams instead of words, which maybe of higher informativeness and distinguishability. We leave it for future work. 

\subsection{Permutation-Invariant Decoding}\label{subsec:decoding}
% 之前的方法使用constrained beam search进行高效解码，它将所有DocID以前缀树or FM index的方式组织起来，每一个docid变成了树上的一个trajectory，
% 从而保证了生成的DocID是valid，即对应于一个存在的文档
% 在TSGen中，term set的所有排列均指向对应文档，naively，这意味着存在N！种可能的解码路径，在一个树中maintain这些序列是intractable的，因此我们实现了permutation-invariant decoding，其能够同时保证validity和optimality。

Constrained beam search is the most prevalent way for generative retrieval to efficiently generate valid DocIDs. 
It creates a prefix tree (a.k.a. trie) encompassing all DocIDs in the corpus and constrains the decoding process along a root-to-leaf trajectory on this trie.
In TSGen, any permutation of the term set DocID leads to the corresponding document. Naively, this yields $N!$ possible trajectories on the trie, which is intractable.
Instead, we propose a permutation-invariant decoding algorithm to replace the constrained beam search. It is characterized by three properties. 
Firstly, it guarantees that the generated DocID is \textit{valid}. In other words, the generation result is precisely one permutation of an existing term set in the corpus. 
Secondly, it allows the model to explore the \textit{optimal} permutation of the term set, so that the model can first generate the terms it reckons more probable, then make fine-grained decisions for others.
Thirdly, it is implemented with an inverted index based structure and hence is competitive in \textit{efficiency}. 
The overall algorithm is depicted in Algotithm~\ref{alg:decoding}. We'll go through the algorithm and give the necessary explanation in the following.

\begin{algorithm}
\caption{Permutation-invariant decoding.}
\label{alg:decoding}
\begin{algorithmic}[1]
\LineComment{Create an empty inverted index that maps each term to documents containing it.}\label{alg:inverted_index_start}
\State $\Phi \gets \mathrm{Inverted\_Index()}$
\For{$D\in\mathcal{D}$}
    \For{$\tau \in \mathcal{T}(D)$}
        \State $\Phi(\tau)$.add(*$D$)
    \EndFor
\EndFor\label{alg:inverted_index_end}
\LineComment{For each beam...}
\State $B\gets\text{Beam Size}$
\For{$1\le b\le B$}\label{alg:initialize_start}
\LineComment{Initialize the generated term with ``<BOS>''.}
\State $X_{b,0}\gets[\text{<BOS>}]$
\LineComment{Initialize the pointers to valid documents.}
\State $Y_{b,0}\gets \{\prescript{*}{}{D}:~D\in\mathcal{D}\}$
\LineComment{Initialize the valid decoding space.}
\State $V_{b,0}\gets \cup\{\mathcal{T}(D):~D\in\mathcal{D}\}$
\EndFor\label{alg:initialize_end}

\LineComment{Start generation...}
\For{$1\le t\le N$}\label{alg:loop_start}
\LineComment{Beam search over the valid decoding space.}
\State $\forall 1\le b\le B,\quad\mathcal{X}_{b,t}\gets X_{b,t-1} \times V_{b,t}$\label{alg:decoding_space}
\State\label{alg:decode_step}
\vspace{-0.4cm}
\begin{align*}
    \quad X_{1,t},\dots,X_{B,t}\gets \underset{X_{b,t} \in \mathcal{X}_{b,t}}{\arg\max}&\sum_{b=1}^B\log p(X_{b,t}\mid X_{b,<t};Q;\Theta) \\
    s.t. X_{i,t}&\ne X_{j,t}
\end{align*}
\LineComment{Reorder valid documents to align with current beams.}\label{alg:reorder_step}
\State $\forall 1\le b\le B,\quad Y_{b,t-1}\gets \mathrm{reorder}(Y_{b,t-1})$
\LineComment{Get documents containing the newly decoded term.}\label{alg:get_doc_step}
\State $\forall 1\le b\le B,\quad x_{b,t}\gets X_{b,t}[-1],\quad y_{b,t}\gets\Phi(x_{b,t})$

\LineComment{Narrow down the valid documents.}\label{alg:narrow_step}
\State $\forall 1\le b\le B,\quad Y_{b,t}\gets Y_{b,t-1}\cap y_{b,t}$

\LineComment{Update the valid decoding space.}\label{alg:get_word_step}
\State $\forall 1\le b\le B,\quad V_{b,t+1}\gets \{\mathcal{T}(D):~D\in Y_{b,t}\} \setminus X_{b,t}$
\EndFor\label{alg:loop_end}

\LineComment{Return the retrieved documents.}
\State\Return{$\{Y_{b,N}\}_{b=1}^B$}

\end{algorithmic}
\end{algorithm}

The beam size is denoted as $B$. We keep track of the following variables:
(1) A list of terms that have been generated until the $t$-th generation step in $b$-th beam, denoted as $X_{b,t}$.
(2) Pointers to documents whose DocIDs contain all the generated terms until the $t$-th step in $b$-th beam, namely the \textit{valid documents}, denoted as $Y_{b,t}$.
(3) The valid decoding space for next-term generation in the $t$-th step and $b$-th beam, denoted as $V_{b,t}$.

In line~\ref{alg:inverted_index_start}-\ref{alg:inverted_index_end}, we build an inverted index to map each term to the document whose DocID contains that term. The document is represented by its pointer for efficiency.
In line~\ref{alg:initialize_start}-\ref{alg:initialize_end}, we initialize the above three variables.
Notably, the initial valid decoding space is all existing terms.
From line~\ref{alg:loop_start}, we start to loop over the $N$ generation steps (since each DocID consists of $N$ terms). 
We first get the valid decoding space by conducting Cartesian product between the already generated terms and all valid terms.
Then we perform a standard beam search over the valid term space, selecting the top $B$ hypotheses based on likelihood. 
Therefore, the generative model is enabled to select any terms in the valid decoding space, which translates to exploring the optimal permutation of the term set on its own.
In line~\ref{alg:reorder_step}, we reorder $Y_{b,t-1}$ so that it aligns with $X_{b,t}$ in terms of the beam index.
Then we look up the inverted index to obtain the documents containing the newly decoded term $X_{b,t}[-1]$, and update the valid documents by intersection in line~\ref{alg:get_doc_step}.
Next, we can get the valid decoding space of the next generation step, which is the union of terms in the valid DocIDs minus the already decoded terms.
At the end of the loop, the valid $B$ documents are returned, which correspond to the generated term sets.

The permutation-equivariant decoding draws the same outcome as the intractable solution based on trie: guaranteeing the validity of the generated term-set DocID while allowing the model to explore the optimal permutation of the term set.
Besides, it is competitive in efficiency as verified in Table~\ref{tab:efficiency}.

\subsection{Iterative Optimization}\label{subsec:optimization}
% 生成式检索要求模型会把所有文档信息（DocID）记在参数中，并且学习查询和DocID之间的相关关系，这calls for finetuning the generative model
% 现有的工作由于DocID预先设定，因此可以直接使用seq2seq learning
% 在TSGen中，term set的所有permutation都指向文档，即有N！种可能的sequence。
% 如何训练模型使得其精准地生成正确的DocID是一个巨大挑战
% 最直接的做法是使用random sample的permutation作为label，并且随时切换。
% 然而，正如在intro里介绍过的，不同的permutation之间亦有区别，有的可能更利于生成，我们希望模型能够沿着其认为最容易生成的permutation，从而得到相关文档的概率能够被最大化。随机sample的permutation并不符合这个要求
% 我们提出了iterative optimization
The generative model to remember the DocIDs in its parameters, meanwhile, it must learn the relevance between the query and the DocIDs. 
This calls for fine-tuning the model with annotated or synthetic data.
Existing works usually employ one or several sequences as DocID, which are defined in advance of the optimization of the model.
Therefore, the Seq2Seq learning can be directly applied, which maps the input query to the relevant DocID.

In TSGen, a set of terms is used as the DocID, and all permutations of the term set lead to the corresponding document. 
This implies $N!$ possible sequences for a single document, while enumerating them is intractable.
A straightforward solution is to randomly sample one or several sequences from these candidates and perform the standard Seq2Seq learning.
However, as mentioned in Section~\ref{sec:intro}, different permutations of the term set may significantly differ in the generation likelihood. 
Thus, a randomly sampled permutation may not be favorable to the model, which means its generation likelihood is low. 
Optimizing the model towards generating in such a permutation may adversely influence its memorization and generalization capability.
Instead, we incentivize the model to follow the permutation that it deems the most probable, which is in line with the goal of our permutation-equivariant decoding algorithm. This is done with our iterative optimization procedure.

In $T$-th iteration ($T$ starts from $1$), we sample the favorable permutation of the term-set DocID from the generative model itself to serve as the learning objective.
Specifically, given the model's latest snapshot $\Theta^{T-1}$ and the query, we let the model generate the permutation it deems the most probable in a similar way to Algorithm~\ref{alg:decoding}. Formally, denote the beam size as $B'$, terms generated until the $t$-th step in the $b$-th beam as $Z_{b,t}$, we perform the following operation on all beams $1\le b\le B'$ and loop over $N$ steps:
\begin{align}
    V'_{b,t}\gets \mathcal{T}(D)\setminus Z_{b,t-1},&\quad\mathcal{Z}_{b,t}\gets Z_{b,t-1} \times V'_{b,t}, \notag \\
    Z_{1,t},\dots,Z_{B',t}\gets \underset{Z_{b,t} \in \mathcal{Z}_{b,t}}{\arg\max}\sum_{b=1}^{B'}&\log p(Z_{b,t}\mid Z_{b,<t};Q;\Theta^{T-1};u), \notag \\
    s.t. Z_{i,t}&\ne Z_{j,t}.
\end{align}
Note that we add a temperature $u$ to the probability so that slightly diversified permutations can be generated.
As a result, we obtain the permutation with high likelihood $\{Z_{i,N}\}_{i=1}^{B'}$ according to model $\Theta^{T-1}$.
Finally, we use query $Q$ as the source, and the result of the first beam ($Z_{1,N}$) as the target of the Seq2Seq learning in this iteration. After convergence, the model is updated as $\Theta^T$.

There are two remaining issues. One is the initial permutation of the term set. Among different options, e.g., purely randomized permutation, or sampling from the pre-trained generative model like T5 and GPT, we empirically find that permuting the terms by their estimated importance in our term selection module brings forth the best performance.
The other one is about the convergence. Although the sampled permutation is always changing, we keep track of the model's retrieval accuracy on a hold-out validation set. 

\vspace{-5pt}
\section{Experiments}

Experiments are conducted to verify the effectiveness (Section~\ref{subsec:effectiveness}), the generalizability (Section~\ref{subsec:generalizability}), the scalability (Section~\ref{subsec:scalability}), and the efficiency (Section~\ref{subsec:efficiency}) of TSGen. 
Meanwhile, we analyse the individual contribution of our technical designs (Section~\ref{subsec:ablation}).

\vspace{-5pt}
\subsection{Settings}
\noindent$\bullet$~~\textbf{Datasets.} 
We leverage two popular datasets that are widely used by previous works on auto-regressive search engines.
One is the NQ320K dataset~\cite{DSI-Tay,NCI-Wang}, which is curated from Natural Questions~\cite{NQ-Kwiatkowski}, containing 109k documents, 320k training queries, 7830 testing queries. 
The other one is MS300K dataset~\cite{Ultron-Zhou,sun2023learning_to_tokenize}, which is curated from MSMARCO~\cite{MSMARCO-Nguyen}, containing 320k documents, 360k training queries, and 772 testing queries.
To investigate the performance of TSGen when scaled to a larger corpus, we also leverage the MSMARCO Passage dataset, which contains 8.8M passages, 500k training queries, and 6980 testing queries.

\noindent$\bullet$~~\textbf{Metrics.}
We measure the retrieval quality at the top-K cut-off by MRR@K (M@K) and Recall@K (R@K), which focus on the perspective of ranking and recall, respectively. 

\noindent$\bullet$~~\textbf{Implementations.}\label{subsec:implementation}
We leverage T5-base~\cite{T5-Raffel} as our backbone. 
We select $N=12$ terms on NQ320K and MS300K, and $N=16$ terms on MSMARCO Passage.
We treat each single word, separated by space, as one term. The term usually contains a single token. Sometimes it contains multiple tokens. We append a ``,'' to the last token, which indicates the termination of the term. 
We can apply this mechanism to other granularities such as n-grams, i.e. a set of n-grams can be used as the DocID. We leave it to future work.
Following the existing works~\cite{NCI-Wang,Ultron-Zhou}, we leverage fine-tuned DocT5~\cite{DocT5Query-Cheriton} to generate synthetic training queries. For each document, we use $10$ synthetic queries on NQ320K and $3$ synthetic queries on MS300K.
When generation, the beam size $B$ is set to 100 throughout the experiments, which is also the same as in previous works.
As for iterative optimization, we perform 2 iterations by default, and we sample the favorable permutation with a temperature $u=3$.
We've uploaded our implementations to \url{https://github.com/namespace-Pt/TSGen}. 

\begin{table*}[t]
  \vspace{-5pt}
  \centering
  \caption{Evaluation of the retrieval effectiveness on NQ320K and MS300K. $\dagger$ denotes the results on NQ320K are copied from~\cite{sun2023learning_to_tokenize}. * indicates significant improvements over the best generative retrieval baseline with p-value < 0.05.}
  \vspace{-10pt}
  \label{tab:main}

  \begin{tabular}{p{0.11\textwidth}p{0.11\textwidth}cccccccccc}
    \toprule
    & & \multicolumn{5}{c}{\textbf{NQ320K}} & \multicolumn{5}{c}{\textbf{MS300K}} \\
    \cmidrule(lr){3-7}
    \cmidrule(lr){8-12}
    \textbf{Category} & \textbf{Method} &  \textbf{M@10} & \textbf{M@100} & \textbf{R@1} & \textbf{R@10} & \textbf{R@100} & \textbf{M@10} & \textbf{M@100} & \textbf{R@1} & \textbf{R@10} & \textbf{R@100} \\
    \midrule
    \multirow{3}{*}{Sparse} & BM25$\dagger$ & -- & 0.211 & 0.151 & 0.325 & 0.505 & 0.313 & 0.325 & 0.196 & 0.591 & 0.861\\
    & UniCOIL & 0.710	& 0.713	& 0.619 & 0.862 & 0.926 & 0.425 & 0.435 & 0.284 & 0.766 & 0.951 \\
    & SPLADEv2 & \underline{0.726} & 0.731 & 0.624 & 0.873 & \textbf{0.954} & 0.443 & 0.452 & 0.328 & 0.779 & \underline{0.956}\\
    \midrule
    \multirow{3}{*}{Dense} & DPR$\dagger$ & -- & 0.599 & 0.502 & 0.777 & 0.909 & 0.424 & 0.433 & 0.271 & 0.764 & 0.948\\
    & ANCE$\dagger$ & -- & 0.602 & 0.502 & 0.785 & 0.914 & 0.451 & 0.455 & 0.299 & \underline{0.785} & 0.953 \\
    & GTR-Base$\dagger$ & -- & 0.662 & 0.560 & 0.844 & 0.937 & \underline{0.484} & \underline{0.485} & \underline{0.332} & \textbf{0.793} & \textbf{0.960} \\
    \midrule
    \multirow{8}{*}{Generative} & DSI & 0.594 & 0.598 & 0.533 & 0.715 & 0.816 & 0.339 & 0.346 & 0.257 & 0.538 & 0.692\\
    & NCI$\dagger$ & -- & 0.731 & 0.659 & 0.852 & 0.924 & 0.408 & 0.417 & 0.301 & 0.643 & 0.851\\
    & GENRE & 0.653 & 0.656 & 0.591 & 0.756 & 0.814 & 0.361 & 0.368 & 0.266 & 0.579 & 0.751 \\
    & Ultron & \underline{0.726} & 0.729 & 0.654 & 0.854 & 0.911 & 0.432 & 0.437 & 0.304 & 0.676 & 0.794\\
    & SEAL$\dagger$ & -- & 0.677 & 0.599 & 0.812 & 0.909 & 0.393 & 0.402 & 0.259 & 0.686 & 0.899\\
    & MINDER & 0.709 & 0.713 & 0.627 & 0.869 & 0.933 & 0.431 & 0.435 & 0.289 & 0.728 & 0.916 \\
    & GenRet$\dagger$ & -- & \underline{0.759} & \underline{0.681} & \underline{0.888} & \underline{0.952} & -- & -- & -- & -- & --\\
    \cmidrule{2-12}
    & \textbf{TSGen} & \textbf{0.771}* & \textbf{0.774} & \textbf{0.708} & \textbf{0.889} & 0.948 & \textbf{0.502}* & \textbf{0.505}* & \textbf{0.384}* & 0.781* & 0.931* \\
    \bottomrule
  \end{tabular}
\end{table*}

\noindent$\bullet$~~\textbf{Baselines.}
We introduce a diverse collection of generative retrieval baselines with different DocID settings: 
DSI~\cite{DSI-Tay}: using hierarchical KMeans IDs; 
NCI~\cite{NCI-Wang}: using layerwise distinguished KMeans IDs. 
GENRE~\cite{GENRE-Cao}: using titles; 
Ultron~\cite{Ultron-Zhou}: using urls; 
SEAL~\cite{SEAL-Bevilacqua}: using all n-grams in the document; 
MINDER~\cite{li2023minder}: using titles, synthetic queries, and n-grams;
GenRet~\cite{sun2023learning_to_tokenize}: using learned DocIDs.
We also compare several traditional retrieval baselines, including the sparse retrieval methods BM25~\cite{BM25-Robertson}, UniCOIL~\cite{UniCOIL-Lin}, and SPLADEv2~\cite{SPLADEv2-Formal}; 
the dense retrieval methods DPR~\cite{DPR-Karpukhin}, ANCE~\cite{ANCE-Xiong}, and GTR~\cite{GTR-Ni}.
Among them, UniCOIL and SPLADEv2 are both based on learned term weights and are closely related to TSGen.

\vspace{-5pt}
\subsection{Main Analysis}\label{subsec:effectiveness}
% 和生成式检索方法对比，效果很好，举两个例子，证明了TSGen的有效性
% 和sequence-based方法对比，其recall更高，这是因为其减少了false pruning的问题，同时mrr更高，这是因为模型可以按照favorable permutation来生成，其拥有更高的likelihood，从而被排在前面
% 和传统方法相比，TSGen在recall上仍有劣势，这是由于即便使用了term set以及相对应的解码算法，生成过程还是会受到false prune问题的影响。这时的false prune和DocID的构成无关，而是源于模型的判断错误，比如一个文档的所有term都被判定为不相关。如何进一步提升模型的准确率，尤其是在大cutoff上，也将成为我们重要的研究课题
We main evaluation results of TSGen on NQ320K and MS300K are reported in Table~\ref{tab:main}. We have the following observations.

Firstly, \textbf{TSGen demonstrates a notable advantage over the strongest generative retrieval baseline on both datasets}. 
For example, on NQ320K, it outperforms GenRet by $+2\%$ on M@100; 
on MS300K, it achieves a relative improvements of $+16\%$ over Ultron on MRR@100.
This verifies the high retrieval quality of TSGen.

Secondly, when compared with natural langauge sequence based DocIDs, e.g. GENRE with titles, Ultron with URLS, and SEAL/MINDER with n-grams, \textbf{TSGen achieves better Recall}. 
This is as expected since the sequence based DocID is more likely to cause the false pruning of the relevant DocID, while TSGen can largely mitigate the problem. 
TSGen also significantly outperforms them on MRR. This is because TSGen can follow the optimal permutation to generate the term-set DocID, which yields a higher likelihood for the relevant DocID so that it is ranked higher.

Thirdly, \textbf{the advantage of TSGen still holds in comparison with traditional retrieval approaches}: It achieves superior ranking performance than all traditional retrieval baselines in terms of MRR and Recall at small cutoffs. This again validates the effectiveness of TSGen.
In our ablation studies, we'll show that the term-set DocID and the permutation-invariant decoding are the main contributors to such advantages. 

Fourthly, despite the above advantages, we may observe that the sparse/dense retrieval baselines may outperform TSGen in terms of R@100 on MS300K.
This is in line with the findings in~\cite{zhou2023reinforce}. 
We conjecture it's because the ``false pruning'' problem, though largely mitigated by TSGen, still occurs.
However, now it does not orient from the DocID and the generation process, instead, it orients from the generative model.
That's to say, the model believes \textit{all terms} in the term-set DocID are irrelevant to the query and hence chooses others for decoding.
How to avoid such errors and learn a better model to make accurate decisions would be left to future work.

\vspace{-5pt}
\subsection{Generalizability Analysis}\label{subsec:generalizability}
% 生成式检索的能力可以被decompose到两个方面。memoryzation即模型记住了训练集中出现的文档的DocID；generalization即对于训练集中没见过的DocID，其能够成功检索。
% 现有的很多方法的generalization能力较差，他们通常使用extensive的数据增强来构造训练数据，从而将所有信息都记住
% 这个是不好的，一方面开销很大，另一方面无法解决添加新文档的问题
% 我们也从这两个维度研究TSGen对比于现有方法的表现，我们选取了一些baseline，分别使用不同的DocID
% -> natural language based方法要好于cluster based，这因为自然语言有泛化性（overlap of terms and similar patterns），因此模型理解了语言，则能成功泛化到没见过的DocID上
% —> TSGen在memorization上较优，但在generalization上显著更优，这是因为我们的Term-Set会给新文档选到之前模型见过的词，并且由于词序任意，模型能够自行找到合适的排序将相关DocID生成，而sequence based方法很容易做错因为模型不熟悉这些新的docid

\begin{table*}
  \vspace{-5pt}
    \centering
    \caption{Evaluation of the memorization and generalization capability on seen and unseen documents, respectively.}
    \label{tab:generalizability}
    \vspace{-10pt}
    \begin{tabular}{p{0.12\textwidth}|cc|cc|cc|cc|cc|cc}
        \toprule
        & \multicolumn{6}{c}{\textbf{NQ320K}} & \multicolumn{6}{|c}{\textbf{MS300K}} \\
        \cmidrule(lr){2-7}
        \cmidrule(lr){8-13}
        & \multicolumn{2}{c|}{\textbf{Seen (50\%)}} & \multicolumn{2}{c|}{\textbf{Unseen (50\%)}} & \multicolumn{2}{c}{\textbf{All (100\%)}} & \multicolumn{2}{|c|}{\textbf{Seen (50\%)}} & \multicolumn{2}{c|}{\textbf{Unseen (50\%)}} & \multicolumn{2}{c}{\textbf{All (100\%)}}\\
        \midrule
        \textbf{Method} & \textbf{M@10} & \textbf{R@10} & \textbf{M@10} & \textbf{R@10} & \textbf{M@10} & \textbf{R@10} & \textbf{M@10} & \textbf{R@10} & \textbf{M@10} & \textbf{R@10} & \textbf{M@10} & \textbf{R@10}\\
        \midrule
        NCI & 0.771 & 0.882 & 0.050 & 0.143 & 0.410 & 0.549 & 0.408 & 0.643 & 0.034 & 0.082 & 0.260 & 0.412 \\
        GENRE & 0.763 & 0.869 & 0.138 & 0.187 & 0.448 & 0.558 & 0.361 & 0.579 & 0.150 & 0.312 & 0.196 & 0.411 \\
        Ultron & \underline{0.782} & {0.891} & {0.300} & {0.383} & {0.471} & {0.570} & \underline{0.432} & {0.676} & {0.197} & {0.246} & {0.313} & {0.492}\\
        MINDER & 0.774 & \textbf{0.907} & \underline{0.303} & \underline{0.415} & \underline{0.488} & \underline{0.639} & {0.431} & \underline{0.728} & \underline{0.285} & \underline{0.433} & \underline{0.335} & \underline{0.569} \\
        \midrule
        \textbf{TSGen} & \textbf{0.809} & \underline{0.900} & \textbf{0.466} & \textbf{0.654} & \textbf{0.552} & \textbf{0.700} & \textbf{0.484} & \textbf{0.766} & \textbf{0.390} & \textbf{0.588} & \textbf{0.391} & \textbf{0.642} \\
        \bottomrule
    \end{tabular}
\end{table*}

\begin{table}[htb]
    \vspace{-5pt}
    \centering
    \caption{Dataset statistics for the evaluation of the memorization and generalization capability of TSGen.}
    \label{tab:dataset}
    \vspace{-10pt}
    \begin{tabular}{lccccc}
        \toprule
        Dataset & Shard & \#Docs & \#Train & \#Validation &\#Test \\
        \midrule
        \multirow{2}{*}{NQ320K} & Seen & 59,739 & 173,447 & 3,000 & 3,915\\
        & Unseen & 50,000 & -- & -- & 3,915\\
        \midrule
        \multirow{2}{*}{MS300K} & Seen & 314,461 & 359,000 & 564 & 772\\
        & Unseen & 289,790 & -- & -- & 439\\
        \bottomrule
    \end{tabular}
\end{table}

The retrieval capability of the generative model can be decomposed into two dimensions. 
1) The \textbf{memorization}, which memorizes the DocID and the query-doc relevance that have been seen during training. 
2) The \textbf{generalization}, which generalizes the learned knowledge to new DocIDs that have not been observed in training.
Many existing works have been shown to have limited generalization capability~\cite{sun2023learning_to_tokenize}. They usually rely on extensive data augmentation (e.g. generate many synthetic queries for each document in the corpus) to alleviate this problem.
However, the problem remains especially when new documents are updated to the corpus~\cite{DSI++-Mehta,CLEVER-Chen}.

In this experiment, we evaluate the memorization and generalization capability of TSGen. 
We partition the corpus into two halves for both NQ320K and MS300K, with training queries preserved for 50\% of the documents (Seen), and with training queries removed for the other 50\% of the documents (Unseen).
The statistics of the curated datasets are reported in Table~\ref{tab:dataset}.
Given the above setting, the generative model is prevented from memorizing any information about the unseen documents during the training stage.
We pick out several strong baselines for comparison, each of which uses a distinct DocID schema. The results are reported in Table~\ref{tab:generalizability}.

According to the results, TSGen marginally outperforms the baselines on the ``seen'' half; nevertheless, its advantage is significantly magnified on the ``unseen'' half.
This indicates the superior generalizability of TSGen.
The reasons are two-fold. 
1) TSGen uses a set of terms as the DocID, which is naturally generalizable across documents thanks to the potential overlap of terms. 
In other words, the unseen documents may share the same term as seen documents. Thus the model learned on the seen half has some prior knowledge of the DocIDs in the unseen half.
Though GENRE, Ultron, and MINDER may also have such prior knowledge thanks thanks to the potentially similar semantics between DocIDs, they may fail to take advantage of it due to the false pruning problem.
2) TSGen adopts the permutation-invariant decoding, which enables the model to explore its favorable permutation of the term set, and hence facilitates the generation of the relevant DocID.

\vspace{-5pt}
\subsection{Scalability Analysis}\label{subsec:scalability}
% scalability->海量文档上的表现
% 以MSMARCO passage为基础
\begin{table}[!t]
  \vspace{-5pt}
    \centering
    \caption{Evaluation of the scalability on MSMARCO Passage~\cite{MSMARCO-Nguyen}. $\dag$ denotes the result copied from~\cite{How_Does_Generative_Retrieval_Scale_To_Millions_Of_Passages-Pradeep}. $\ddag$ denotes the result copied from~\cite{li2023minder}.}
    \label{tab:scalability}
    \vspace{-10pt}
    \begin{tabular}{p{0.2\textwidth} c c}
    \toprule
    \textbf{Model} & \textbf{\#Training Queries} & \textbf{M@10} \\
    \midrule
    BM25 & -- & 0.187 \\
    DPR & 0.5M & \textbf{0.314} \\
    \midrule
    DSI+DocT5$\times$1 & 8.8M & 0.075 \\
    DSI+DocT5$\times$40$\dag$ & 352M & 0.133 \\
    MINDER$\ddag$ & 7.2M & 0.186 \\
    TSGen & 8.8M & 0.195 \\
    \bottomrule
    \end{tabular}
\end{table}

Scalability is a crucial challenge for generative retrieval and is of wide interest to the community~\cite{How_Does_Generative_Retrieval_Scale_To_Millions_Of_Passages-Pradeep}.
In this experiment, We evaluate TSGen on the entire MSMARCO Passage dataset, which contains 8.8M passages. 
Following the setting in~\cite{How_Does_Generative_Retrieval_Scale_To_Millions_Of_Passages-Pradeep}, we augment each document with $M$ synthetic queries generated by a DocT5~\cite{DocT5Query-Cheriton} model, and only use the synthetic queries as the training data.
Due to limited computation resources, we set $M=1$ (the original work set $M=40$), resulting in $8.8M$ training queries.
We set the number of selected terms per passage (i.e. $N$) to $16$. This results in 48,577 term-set collisions. We examine some cases and find most collisions are coming from different editions of the same page (the contents of two passages are almost identical). 
In this case, knowing that the computation cost grows as $N$ increases, we do not further scale $N$. Instead, we keep uniqueness by applying simple heuristics given $N$ fixed to 16. Concretely, if two documents' term sets collide, we replace one document's last selected term with another less weighted term in it to distinguish those two documents.

The results are reported in Table~\ref{tab:scalability}. 
First of all, TSGen is advantageous against DSI on such a massive corpus: when trained with the same amount of data (8.8M training queries), TSGen significantly improves the MRR of DSI. 
It even outperforms DSI scaled with 40 augmented queries, which consumes 40 times more training data and hence is much more expensive.
Besides, TSGen improves the strong baseline MINDER by 5\% when trained with a similar amount of queries. 
These observations validate the scalability of TSGen. 
Lastly, generative retrieval methods still lag far behind the performance of dense retrieval on such a large corpus. This has been a widely-known issue of generative retrieval methods.
Several concurrent works point out that utilizing contrastive learning~\cite{zeng2023scalable_and_effective,li2023learning_to_rank_in_generative_retrieval} in training may improve the effectiveness on a large corpus, which may be combined with TSGen for further improvements.

\vspace{-5pt}
\subsection{Efficiency Analysis}\label{subsec:efficiency}
\begin{table}[!t]
  \vspace{-5pt}
  \centering
  \caption{Evaluation of the efficiency on NQ320K. }
  \label{tab:efficiency}
  \vspace{-10pt}
  \begin{tabular}{p{0.09\textwidth}ccc}
  \toprule
  \textbf{Method} & \textbf{Memory} & \multicolumn{2}{c}{\textbf{Query Latency (s)}}\\
  \cline{3-4}
  & (MB) & beam size = 10 &  beam size = 100\\
  \midrule
  DSI & 12 & 0.03 & 0.21 \\
  NCI & 12 & 0.03 & 0.21 \\
  GENRE & 27 & 0.05 & 0.47 \\ 
  Ultron & 32 & 0.08 & 0.64 \\ 
  MINDER & 210 & 0.32 & 3.14 \\
  \midrule
  \textbf{TSGen} & 35 & 0.06 & 0.69\\
  \bottomrule
  \end{tabular}
\end{table}
The running efficiency is evaluated in Table~\ref{tab:efficiency}. Particularly, we measure the memory consumption for hosting the DocIDs of the entire corpus; we also measure the time cost (query latency) with different beam sizes. 
DSI and NCI enjoy the smallest memory usage and the lowest query latency thanks to their short DocID (only 10 integers).
GENRE and Ultron require more space and are a little slower than DSI, because their DocID sequences are longer.
% All four approaches use a trie to constrain the beam search. 
Our method, TSGen, leverages an inverted index to perform permutation-invariant decoding. It achieves similar efficiency as Ultron.
MINDER employs an FM index, which consumes much more memory and is also slower than all other approaches.

\vspace{-5pt}
\subsection{Ablation Studies}\label{subsec:ablation}
\begin{table}[t]
    \centering
    \caption{Ablation studies on NQ320K. The default settings of TSGen are marked with \textit{*}.}
    \label{tab:ablation}
    \vspace{-10pt}
    \begin{tabular}{p{1.8cm}lccccc}
        \toprule
        \textbf{Factor} & \textbf{Setting} & \textbf{M@10}& \textbf{R@10} & \textbf{R@100} \\
        \midrule
        \multirow{2}{*}{DocID}
        & Sequence & 0.749 & 0.864 & 0.921 \\
        & {Term-Set}$^*$ & \textbf{0.771} & \textbf{0.889} & \textbf{0.948}\\
        \midrule
        \multirow{3}{2cm}{Term\\Selection}
        & Random & 0.628 & 0.739 & 0.811\\
        & Title & 0.743 & 0.856 & 0.915\\
        & {Learned}$^*$ & \textbf{0.771} & \textbf{0.889} & \textbf{0.948}\\
        \midrule
        \multirow{2}{2cm}{Optimization} & Non-Iterative & 0.751 & 0.868 & 0.932\\
        & {Iterative}$^*$ & \textbf{0.771} & \textbf{0.889} & \textbf{0.948}\\
        \midrule
        \multirow{3}{*}{Term Number} & 8 & 0.760 & 0.879 & 0.940 \\
        & 16 & \textbf{0.771} & \textbf{0.889} & 0.947 \\
        & 12* & \textbf{0.771} & \textbf{0.889} & \textbf{0.948}\\
        \midrule
        \multirow{3}{2cm}{Initial\\Permutation} & Random & 0.728 & 0.857 & 0.931\\
        & Likelihood & 0.716 & 0.847 & 0.915\\
        & {Weight}$^*$ & \textbf{0.771} & \textbf{0.889} & \textbf{0.948}\\
        \midrule
        \multirow{4}{2cm}{Synthetic\\Queries (SQ)} 
        & Ultron w.o. SQ & 0.670 & 0.779 & 0.845\\
        & NCI w.o. SQ & -- & 0.679 & 0.909\\
        % \midrule
        & TSGen w.o. SQ & 0.728 & 0.846 & 0.925 \\
        & {TSGen}$^*$ & \textbf{0.771} & \textbf{0.889} & \textbf{0.948} \\ 
        \midrule 
        \multirow{5}{2cm}{Model\\Scale}
        & DSI large & 0.613 & 0.733 & 0.835 \\
        & NCI large & -- & 0.885 & 0.945 \\
        & GENRE large & 0.663 & 0.770 & 0.828 \\
        & SEAL large & -- & 0.812 & 0.909 \\
        & {TSGen large} & \textbf{0.779} & \textbf{0.896} & \textbf{0.954} \\
        \midrule
        \multirow{3}{2cm}{Beam\\Size} & 10 & 0.770 & 0.876 & -- \\
        & 200 &  \textbf{0.771} & \textbf{0.892} & \textbf{0.951}\\
        & 100* &  \textbf{0.771} & {0.889} & {0.948} \\
        \bottomrule
    \end{tabular}
\end{table}

The ablation studies are performed for each influential factor in TSGen based on NQ320K dataset as Table~\ref{tab:ablation}.  

\noindent$\bullet$~~\textbf{DocID.} 
We compare the proposed term-set DocID with the sequence based one. 
For the latter, the terms are ordered as a sequence by their estimated weights (empirically more competitive than other sequence orders). 
It can be observed that the retrieval quality of term-set DocID is notably superior to that of the sequence DocID. 
As discussed, the term-set DocID together with the permutation-invariant decoding effectively mitigates the false pruning problem, which often happens on the sequence based DocID.

\noindent$\bullet$~~\textbf{Term Selection.}
We compare our learned term selection with two alternatives: 
Random, the randomly selected terms from the document; 
Title: terms within the title. 
Firstly, there are huge differences between different term selection strategies, which verifies the importance of term selection. 
Secondly, although directly making use of title is a strong baseline (also a common practice in many works~\cite{GENRE-Cao,de-cao-etal-2022-multilingual,Ultron-Zhou}), our learned selection strategy is more effective.  
Specifically, the term weights learned from relevance capture the relationship between queries and documents.
Thus, terms selected based on these weights can distinguish the document from others, which are suitable to serve as the DocID.

\noindent$\bullet$~~\textbf{Term Number.} We compare three settings of the number of selected terms for each document i.e. $N$. It can be observed that selecting less terms for each document degrades the ranking performance more than the recall. 
This can be attributed to the fact that the identifier collisions are more likely to happen, and hence TSGen does not know how to rank documents if they share the same DocID, yet including them all in the retrieval results can roughly maintain the the Recall@100 performance. 
On the other hand, selecting more terms for each document cannot further improve the retrieval quality. 
Therefore, our selecting protocol uses the smallest $N$ while keeping discrimination, i.e. 12 on NQ320K.

\noindent$\bullet$~~\textbf{Optimization.}
We compare our iterative optimization with its non-iterative variant where the DocID's permutation is fixed as its initialization. 
Note that other settings (e.g. permutation-invariant decoding) are kept the same. 
It can be observed that our proposed optimization approach indeed contributes to retrieval quality. 
Such an advantage is easy to comprehend, considering that the training objective (the permutation of the term-set DocID) can be iteratively updated to keep consistent with the goal of our permutation-invariant decoding in the testing stage. 

\noindent$\bullet$~~\textbf{Initial Permutation.}
We compare three approaches for initializing the permutation in the first iteration. 
1) Random: the selected terms are randomly permuted; 
2) Likelihood: the selected terms are permuted by the generation likelihood of the pre-trained T5; 
3) Weight: the selected terms are permuted by their estimated weight in the term selection module. 
We can observe that the initialization turns out to be another critical factor for TSGen: the importance-based method is notably stronger than the other two baselines. 
This is probably because the importance-based initialization is easier to generate and better reflects the query-document relationship. 

\noindent$\bullet$~~\textbf{Synthetic Queries.} 
Augmenting synthetic queries to is a widely used strategy to improve generative retrieval~\cite{NCI-Wang,Ultron-Zhou,DSI++-Mehta}. 
It is also helpful for TSGen.
Specifically, TSGen's retrieval quality is substantially improved on top of query generation. 
Besides, the relative improvement of TSGen is mainly from the proposed term-set DocID and its generation workflow, rather than the extra data augmentation. 
When query generation is disabled, TSGen maintains its advantage over other baselines. 

\noindent$\bullet$~~\textbf{Model Scale.} The scaling-up of the backbone generative model is another common approach to enhance generative retrieval. In our experiment, empirical improvements are also observed when we switch to a T5-large backbone. 
Meanwhile, it maintains the advantage when other baselines are scaled up as well.

\noindent$\bullet$~~\textbf{Beam Size.} We default to use 100 as the beam size following previous works, and juxtapose the other two different settings. 
It can be observed that the Recall@10 drops when decreasing the beam size to 10. This is as expected since a smaller beam size keeps fewer candidates in each generation step and hence terms in the relevant DocID may not be included. 
Another interesting observation is that the MRR@10 almost stays the same when decreasing or increasing the beam size. 
This is because TSGen can generate the relevant DocID in its favorable permutation, thus assigning a high likelihood to it and ranking it at the top, regardless of changing the beam size. 

\vspace{-5pt}
\section{Case Study}
\begin{table}[!tb]
    \centering
    \small
    \caption{Case study of TSGen on MS300K. The displayed document is the target of both query 1 and query 2. The term set DocID is ordered by learned term weights. Terms matching with the query are highlighted in \highlight{yellow}.}
    \vspace{-10pt}
    \label{tab:case}
    \begin{tabular}{l p{0.35\textwidth}}
    \toprule
        \multicolumn{2}{c}{\textbf{Document}} \\
        \textbf{Title: } & What Foods Not to Eat When Having High Creatinine \\
        \hdashline
        \textbf{URL: } & http://www.kidney-treatment.org/creatinine/162.html \\
        \hdashline
        \textbf{Body: } & ... Creatinine is a breakdown product of creatine. Since kidneys are responsible for discharging creatinine in blood, so when kidneys are injured for some reason, creatinine level in blood increases. Therefore, high creatinine level indicates there are lots of wastes in blood. 
        What foods not to eat when having high creatinine? 1. Adjust protein intake...\\
        \hdashline
        \textbf{Term Set: } & kidney, foods, eat, creatinine, creatine, blood, high, level, snack, salted, wastes, potassium\\
        \midrule
        \midrule
        \textbf{Query 1: } & What \highlight{food} should not be \highlight{eaten} in \highlight{kidney} failure\\
        \hdashline
        \textbf{Permut. 1: } & \highlight{kidney}, \highlight{foods}, \highlight{eat}, level, high, salted, snack, creatinine, creatine, blood, potassium, wastes\\
        \midrule
        \textbf{Query 2: } & \highlight{foods} to raise my \highlight{creatinine} \highlight{level}\\
        \hdashline
        \textbf{Permut. 2: } & \highlight{creatinine}, \highlight{creatine}, \highlight{foods}, high, \highlight{level}, eat, kedney, blood, mean, snack, wastes, salted, potassium \\
    \bottomrule
    \end{tabular}
\end{table}
In Table~\ref{tab:case}, we show an example on MS300K to qualitatively evaluate TSGen. Compared with the title or URL, the term-set DocID effectively summarize the information of the document. Besides, TSGen generates the same term set in its favorable permutation for different queries. It puts the term that matches with the query at the front so that the likelihood is maximized.

\section{Conclusion and Future Work}

In this work, we present TSGen, a novel framework for generative retrieval to mitigate the false pruning problem stemming from the natural language sequence DocID. 
It employs a set of terms as the DocID instead of one or several sequences.
On top of the term-set DocID, we propose the permutation-invariant decoding, with which any permutations of the term set will always lead to the corresponding document. 
We further devise an iterative optimization procedure to incentivize the model to generate the relevant term set in its favorable permutation.
With comprehensive experiments, we empirically verify that TSGen is more effective, generalizable, and scalable than existing generative retrieval methods with competitive efficiency.
In the future, we would like to explore other granularities to form the DocID, for example, the n-gram set.
Besides, we may combine the recent contrastive learning techniques to further boost the performance on a large corpus.

\section*{Acknowledgement}
This work was supported by the Beijing Natural Science Foundation No. L233008, the National Natural Science Foundation of China No. 62272467,  the fund for building world-class universities (disciplines) of Renmin University of China, and Public Computing Cloud, Renmin University of China. The work was partially done at the Engineering Research Center of Next-Generation Intelligent Search and Recommendation, MOE.

% \begin{acks}
% \end{acks}

\clearpage
\bibliographystyle{ACM-Reference-Format}
\balance
\bibliography{sample-base}

\end{document}